\documentclass[fleqn,10pt]{wlscirep}
\usepackage{amsmath}

\title{Topologically protected localised states in spin chains}

\author[1,*]{Marta P. Estarellas}
\author[1]{Irene D'Amico}
\author[1]{Timothy P. Spiller}
\affil[1]{Department of Physics, University of York,
		York YO10 5DD, United Kingdom}

\affil[*]{mpee500@york.ac.uk}

\begin{abstract}
We consider spin chain families inspired by the Su, Schrieffer and Hegger (SSH) model.  We demonstrate explicitly the topologically induced spatial localisation of quantum states in our systems. We present detailed investigations of the effects of random noise, showing that these topologically protected states are very robust against this type of perturbation. Systems with such topological robustness are clearly good candidates for quantum information tasks and we discuss some potential applications. Thus, we present interesting spin chain models which show promising applications for quantum devices.
\end{abstract}
\begin{document}

\flushbottom
\maketitle
%
%
\thispagestyle{empty}

\section*{Introduction}

The topological confinement of quantum states has engaged the condensed matter community for a few decades \cite{jakiw1976,jakiw1981}. Now this field is receiving increasing interest due to its potential applications for topological quantum computation \cite{sarma2015}, quantum state transfer \cite{yao2013} and quantum memories \cite{kornschelle2013,pedrocchi2015}. Such systems can embody quantum information non-locally, leading to topologically protected states and opening new paths in the search for a robust quantum processing architecture \cite{elliott2015}. Further interest is also growing in applications with spin chains \cite{saket2010,saket2013,levitov2001,srivinasa2007}.
	
	The Su, Schrieffer and Hegger (SSH) model  was first presented in 1979, to describe soliton formation in polyacetylene \cite{SuSchriefferHeeger}. 
In this model the presence of the phonon-electron interaction  causes the distortion (dimerization) of the lattice and the consequent possible formation of `topological solitons' \cite{jakiw1981}. Topological solitons occur in systems with degenerate ground states and represent boundaries between domains with different ground states. Because of this 
topological nature, in the simplest picture of a single soliton (the SSH chain), this causes an electronic (spin in the case here examined) excitation at zero energy. This excitation is  
spatially localised over the symmetry breaker (the soliton). Importantly this excitation is protected by a substantial energy gap.
Recently, an analogy of this model has been implemented with a set of identical, coupled dielectric resonators placed in a microwave cavity \cite{poli2014}, inducing spatially confined states. The presence of these topological solitons in one-dimensional systems can be understood and explained through calculation of the Zak phase \cite{zak1989}. This non-vanishing phase arises in a periodic parameter space and is different for topologically distinct configurations of the system. The value of the phase itself is gauge dependent; however, crucially, phase differences are not and these enable topological classification of configurations. At interfaces between topologically distinct configurations  (i.e. configurations with different Zak phases), protected and localised states arise, localised at the interface. This topological characterization and analysis of SSH-like models has been widely studied in previous works \cite{bloch2013,schomerus2013,poli2014,delplace2011,bello2016}. We will discuss this analysis in the context of our work later in the paper.

In this paper we analyse families of finite linear spin chains, which have topological solitons embedded into the structure of their spin-spin coupling terms. These types of system are inspired by the SSH chain, and the spin model presented here can be related to the original fermionic SSH model through the (non-local) Jordan-Wigner transformations that relate the spin and fermion operators \cite{parkinson2010,jordanwigner1993}. When configurations with more than one soliton -- and with a more complicated symmetry breaking pattern -- are explored, more than one localised state is found and, importantly, all of these localised states are protected by energy gaps.
Again, these energy gaps and corresponding protected states exist because of the topology of the system. We consider in detail two different families of symmetric configurations and give a detailed study of the robustness of these chains against noise. 

In recent years spin chains have acquired growing importance within the field of quantum information processing, as a means of efficiently transferring information \cite{bose2007,kay2010}, and for creating and distributing entanglement \cite{spiller2007}. Spin chains are of particular interest due to their versatility to be engineered e.g, to allow for ``perfect state transfer'' \cite{christandl2004,chris2005,plenio2004} across the chain. By tuning the couplings, here we show how we can construct our chains to be topologically analogous to the SSH model. Experimentally, this can be done for any system where it is possible to engineer the couplings between the sites, e.g., electrons and excitons trapped in nanostructures \cite{damico2007,damico2006,niko2004}, nanometer scale magnetic particles \cite{tejada2001} or a string of fullerenes \cite{twamley2003}. More specific hardware in which these SSH-like systems have been engineered are, for example, edges of graphene ribbons \cite{delplace2011}, edges of honeycomb arrays of microcavity pillars \cite{jacqmin2014}, optical waveguides \cite{Redondo2016} or Bose-Einstein condensates of $Rb^{87}$ atoms in suitable optical lattice potentials \cite{meier2016,bloch2013}. In fact, as the polyacetylene-inspired chains studied here require only two different couplings, there is potential for relative ease of chemical engineering of such systems. 

As already mentioned, the localised states within the chains we analyse are of topological origin and are within energy gaps. From a quantum information perspective, these energy gaps are -- per se -- offering protection against environmental effects to these energy-isolated and localised states, which is certainly advantageous when using these states to encode quantum information, 
even just for quantum memory.   
One surprising result of our study is that we find that the zero energy states are protected from noise even when this is comparable and {\it even larger} than the 
energy gap.
	
	In the following, we study in detail the robustness of topologically localised states, by exploring spin chains with distinct coupling patterns. We first explain the physical model for our spin chain. We analyse the eigenstate spectrum for two different spin chain families, each in the one-excitation subspace. Clearly, any physical implementations will always be subject to errors and imperfections due to the presence of field fluctuations and fabrication defects. We therefore investigate in detail the effects of introducing random noise. As will be seen, the localised states persist with high fidelity. We conclude by considering some applications of the effects we observed, as a tool to manipulate the properties and behaviour of spin chains as quantum devices.

\section*{The Model}

\begin{figure}
		\centering
		\includegraphics[scale=0.6]{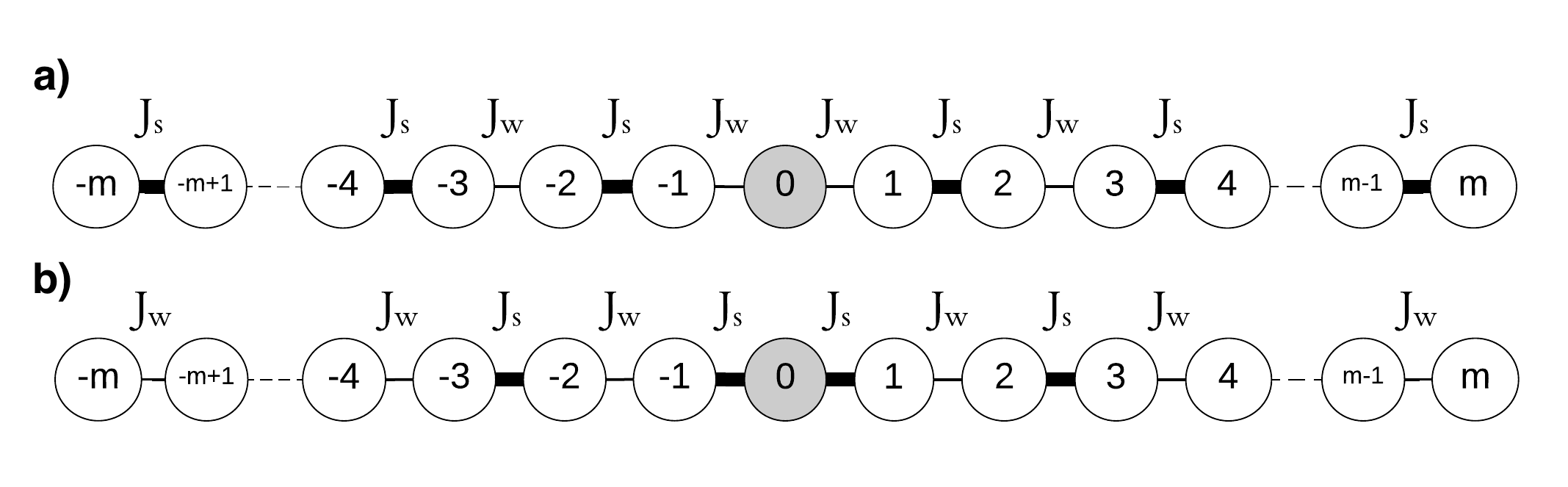}
		\caption{Spin chains, for a chain with site 0 (a) weakly coupled and (b) strongly coupled.  }
		\label{fig1}
	\end{figure}

Analogous to the polyacetylene single and double bonds in the original SSH model, here we consider symmetric chains with an odd number of spins with alternating weak ($J_{w}$) and strong ($J_{s}$) couplings. Each chain is symmetric with respect to the mid point, here labelled as site 0. The coupling nature of this site 0 spin determines the coupling pattern of the rest of the chain (Fig.\ref{fig1}). 
	
	The time independent Hamiltonian of an $N$-site spin chain (where $m=(N-1)/2$) is given by,
	
	\begin{eqnarray}
        \centering
		\label{hami}
		&{\cal{H}} = \sum_{i=-m}^{m}\epsilon_{i}|1\rangle \langle 1|_{i} + \sum_{i=-m}^{m-1} J_{i,i+1}[ |1\rangle \langle 0|_{i} \otimes |0\rangle \langle 1|_{i+1} + |0\rangle \langle 1|_{i} \otimes |1\rangle \langle 0|_{i+1}]
	\end{eqnarray}	
	
	In this work we consider the single excitation energies $\epsilon_{i}$ to be independent of the site $i$ (and so zero for convenience), until later when diagonal disorder is added. At any site, a single excitation $|1\rangle$ symbolises an ``up'' spin, in a system that is initially prepared to have all sites in the spin ``down'' $|0\rangle$ state. The coupling strengths $J_{i,i+1}$ between two nearest-neighbour sites $i$ and $i+1$ are pre-engineered depending on the type of chain used. In the examples presented here, we will demonstrate strongly localised states, non-overlapping protected states with good energy gap protection. To this end we choose a ratio between the couplings of $J_{s}/J_{w}=40$, with a chain length of $N=101$. With such a coupling ratio, the exponential decay of the localised amplitudes \cite{poli2014} is such that our states are indeed highly localised and non-overlapping. However we note that for the chosen coupling ratio example, as long as $N$ exceeds in case (a) 5 -- the shortest limit allowed for this chain type -- and in case (b) approximately 45, the localised state profile does not change and, for case (b), where more than one localised states is present, the localised states have negligible overlap. Thus our example of $N$=101 is well into a region where the localised states are `$N$-independent'. However, we stress that, even for coupling ratios an order of magnitude smaller than 40, the localisation of states associated to solitons is still significant. If some overlap between different localised states is required, this can be achieved by either reducing the chain length or reducing the size of the coupling ratio. We shall give an example later in the paper, achieving this overlap through use of a shorter chain in order to achieve perfect state transfer.
	
	For convenience, from now on we define $J_{s} \equiv \Delta$ and $J_{w} \equiv \delta$. 
	
	For the first case (Fig.\ref{fig1}-a), site 0 is weakly coupled ($J_{-1,0} = J_{0,1}=\delta$) to the rest of the chain such that,  
	
	\begin{equation}
    \centering
	i\in2\mathbb{Z}
	\left\{%
	\begin{array}{ccc}
	J_{i,i+1}\equiv \delta,&J_{i+1,i+2}\equiv \Delta &  (0\leq i< m)\\
	J_{i,i+1}\equiv \Delta,&J_{i+1,i+2}\equiv \delta & (-m\leq i< 0) 
	\end{array}
	\right.
	\end{equation}
	
	In the second case (Fig.\ref{fig1}-b), site 0 is strongly coupled to the rest of the chain, $J_{-1,0} = J_{0,1}=\Delta$,
	
	\begin{equation}
	i\in2\mathbb{Z}
	\left\{%
	\begin{array}{ccc}
	J_{i,i+1}\equiv \Delta,&J_{i+1,i+2}\equiv \delta &  (0\leq i< m)\\
	J_{i,i+1}\equiv \delta,&J_{i+1,i+2}\equiv \Delta & (-m\leq i< 0) 
	\end{array}
	\right.
	\end{equation}

\section*{Protected states: Spatial Localisation and Energy Gaps}

\begin{figure*}
		\centering
		\includegraphics[scale=0.32]{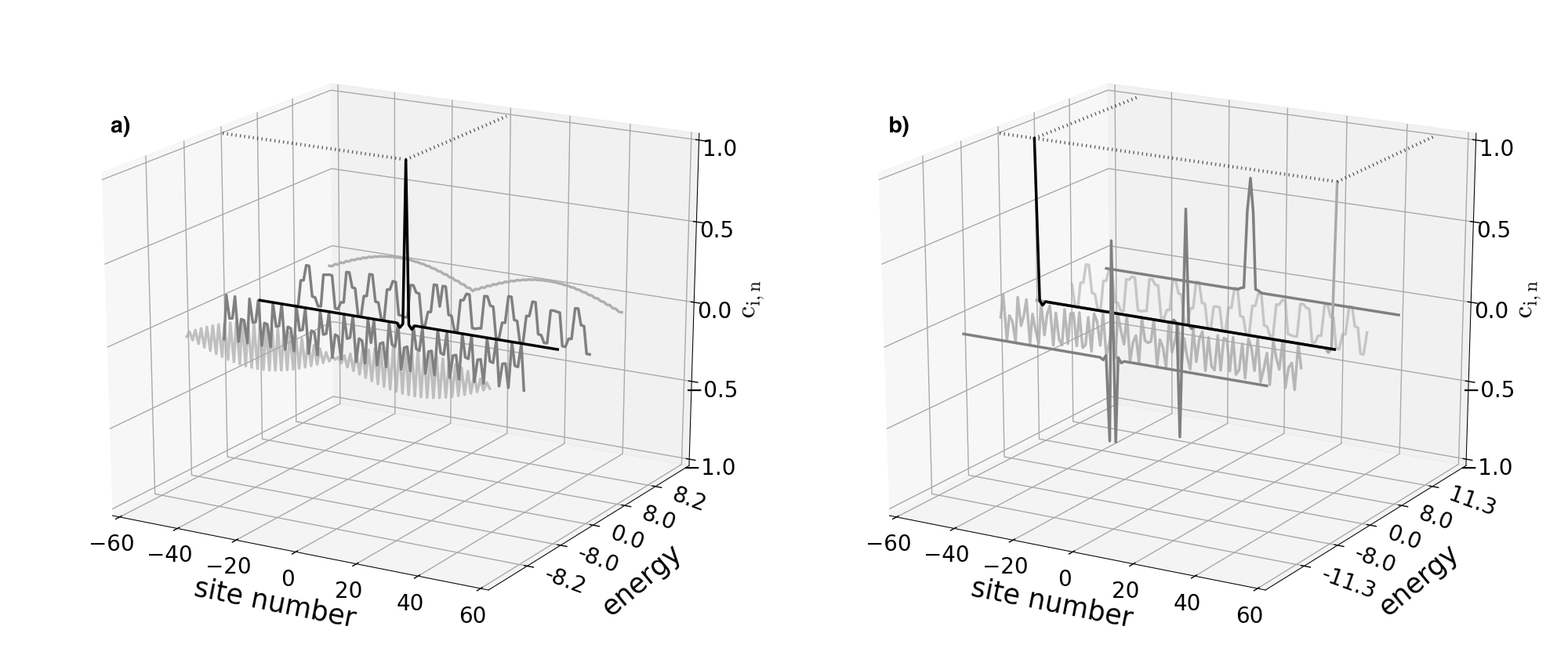}
		\caption{State amplitudes versus site number $i$ and energy for relevant examples of eigenstates, for $N$=101 site spin chains with site 0 (a) weakly coupled, and (b) strongly coupled. For clarity, the y-axis representing the energy has some band gap regions omitted; hence it is not to scale. In case (a) we observe that the localised state peaks at $\sim 1$ (0.9988), but even with a much lower ratio ($J_{s}/J_{w}=4$) it still peaks at 0.8824. For case (a), in addition to the zero energy localised state, we show four extended eigenstates, corresponding to that at the bottom of the lower band (see Fig. \ref{energy}) with energy $-\Delta-\delta=-8.2$, that in the middle of the lower band with energy $-\Delta=-8$, that in the middle of the upper band with energy $\Delta=8$, and that at the top of the upper band with energy $\Delta+\delta=8.2$. For case (b), we show the three zero energy localised states, along with the localised states band with energy $-\sqrt{2}\Delta=-11.3$ below the lower, and with energy $\sqrt{2}\Delta=11.3$ above the upper band. Also shown are two examples of extended states, those in the middle of the lower and upper bands, with energies of $-\Delta=-8$ and $\Delta=8$, respectively}
		\label{eigen}
	\end{figure*}

\begin{figure*}[ht]
		\centering
		\includegraphics[scale=0.35]{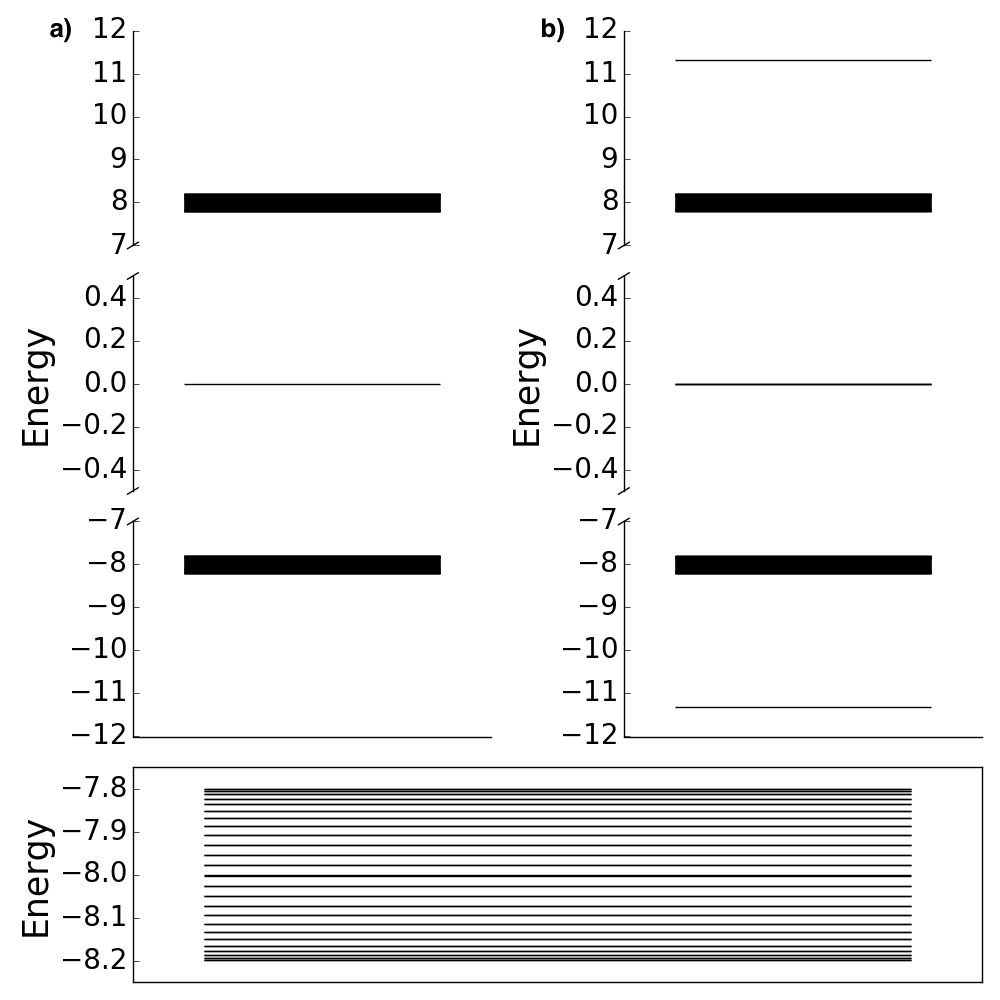}
		\caption{Energy spectrum of $N$=101 spin chains with site 0 (a) weakly, and (b) strongly coupled. The bottom panel represents a zoom of the lower energy band for case (a). In both cases each band state is two-fold degenerate. In case (a) one single localised state sits in the middle, and case (b) three degenerate states sit at zero energy, with a further state above the upper band and one below the lower band. }
		\label{energy}
	\end{figure*}

We now study the single-excitation eigenstates and band structures of these two families of chains. The study of the eigenstates, $|\varphi_n\rangle$, of the system shows localisation signatures for both families of chains studied here. The left panel of Fig.\ref{eigen} shows the amplitudes as a function of site number $i$, $c_{i,n}=\langle i|\varphi_n \rangle$  and energy for the relevant eigenstates, of which there are $N$. It demonstrates that at site 0 of a weakly coupled 101 chain (type (a)), one of the states (black profile) peaks at unity (0.999) with energy zero. Of the other 100 states, $|\varphi_{n=1}\rangle$, $|\varphi_{n=25}\rangle$, $|\varphi_{n=75}\rangle$, $|\varphi_{n=101}\rangle$, with energies $-\Delta-\delta$, $-\Delta$, $\Delta$, $\Delta+\delta$ respectively, are shown (the extremal band states, see Fig. \ref{energy}). These illustrate how the rest of the eigenstates are delocalised along the chain. The right panel of Fig.\ref{eigen} shows the amplitudes for relevant states of case (b). There are three zero-energy degenerate localised eigenstates: two of them peak at unity (0.999) at the two end sites and the other peaks at $|0.499|$ at sites $i=-1$ and $i=1$ (with zero amplitude at site $i=0$). Two further states localised at the chain centre,$|\varphi_{n=1}\rangle$ and $|\varphi_{n=101}\rangle$, have energies $-\sqrt{2}\Delta$ and $\sqrt{2}\Delta$, respectively. The remaining eigenstates are delocalised along the chain, as illustrated with the two states at energies $\pm\Delta$.
	
	We show the full energy spectra for both chains in Fig. \ref{energy}. Each energy spectrum clearly demonstrates the presence of states protected by energy gaps; these correspond to the localised eigenstates shown in Fig. \ref{eigen}. In case (a), Fig.\ref{energy}-a, the system presents one eigenstate in the middle of the gap between twenty-five higher and twenty-five lower twofold degenerate states. In case (b), Fig.\ref{energy}-b, displays instead three degenerate eigenstates in the middle of the gap, corresponding to the ones peaking at the beginning, neighbour sites of the center and end of the chain. These states are surrounded by twenty-four higher/lower twofold degenerate states and one highest/lowest state forming two other bands against the bulk of states. A further two states occur at energies $\pm2\sqrt{\Delta}$; these are the additional states peaking in the chain centre in Fig.\ref{eigen}, also protected by energy gaps.
	
	The spectra and eigenstates of these spin chains can be understood using the ingredients of (i) symmetry and (ii) dimer states.
	For (i) we note that the operator $\cal{M}$ that reflects the system about its middle site ($i=0$) commutes with the Hamiltonian and so non-degenerate eigenstates of the system must each be even or odd (with eigenvalue $+1$ or $-1$) under $\cal{M}$. 
	For (ii), given the dominance of $\Delta \gg \delta$, it is useful to view the pairs of sites coupled strongly as dimers. For two such strongly coupled sites $|i\rangle$ and $|i+1\rangle$ the dimer eigenstates are 
	$\frac{1}{\sqrt{2}}\left(|i\rangle \pm |i+1\rangle \right)$ with eigenvalues $\pm \Delta$. Using these dimer states (rather than the site states) as basis states explains the positions of the energy bands in Fig. 3 for both cases (a) and (b). The mid points of these bands are at the dimer energies $\pm \Delta$. To understand the spectra and eigenstates in more detail we discuss cases (a) and (b) separately.
	
	Case (a): Here we have $m/2$ weakly coupled dimers either side of and weakly coupled to a central site state $|i=0\rangle$. Using these states as a basis, the higher energy ($+\Delta$) dimer states for positive $i$ can be treated as an $m/2$-site weakly coupled chain with a constant coupling of $\delta/2$ \cite{bose2003}. The eigenstates are therefore delocalised superpositions of the dimer states with eigenvalues forming a band from 
	$\Delta-\delta$ to $\Delta+\delta$, with a characteristic cosine distribution within the band \cite{bose2003}, as seen in the bottom panel of Fig.\ref{energy}. The higher energy ($+\Delta$) dimer states for negative $i$ can be treated as a similar 
	$m/2$-site weakly coupled chain, with eigenvalues degenerate with the positive $i$ band. Enforcing the symmetry under reflection $\cal{M}$ means that the final upper band eigenstates for case (a) are even or odd superpositions of the positive and negative $i$ band states. On an energy scale of $\delta$, these odd and even states are degenerate, which explains why each band level, as shown for example in Fig. 3, is doubly degenerate. Clearly a similar analysis follows for the positive and negative $i$ lower energy 
	($-\Delta$) dimer states, giving a band of doubly degenerate states ranging from $-\Delta-\delta$ to $-\Delta+\delta$, with the same characteristic cosine distribution. The full set of $2m+1$ eigenstates and eigenvalues for case (a) is completed with the inclusion of the localised state $|i=0\rangle$ that has zero energy.
	
	Case (b): Here we have end sites ($|-m\rangle$ and $|m\rangle$) each weakly coupled to systems of $(m-2)/2$ dimers which each weakly couple to a trimer (comprising sites $|-1\rangle$, $|0\rangle$ and $|+1\rangle$) in the middle. The trimer states are given in Eq.\ref{eigenstates}, with $|\phi_{-}\rangle$ having energy 
	$-\sqrt{2} \Delta$, $|\phi_{0}\rangle$ having energy zero and $|\phi_{+}\rangle$ having energy $\sqrt{2} \Delta$. 
	
	\begin{equation}
	|\phi_{-}\rangle=
	\left(
	\begin{matrix}
	-1/2\\1/\sqrt{2}\\-1/2
	\end{matrix}
	\right)
	|\phi_{0}\rangle=
	\left(
	\begin{matrix}
	-1/\sqrt{2}\\
	0\\
	1/\sqrt{2}\\
	\end{matrix}
	\right)
	|\phi_{+}\rangle=
	\left(
	\begin{matrix}
	1/2\\
	1/\sqrt{2}\\
	1/2\\
	\end{matrix}
	\right)
	\label{eigenstates}
	\end{equation}
	
	The five localised states in case (b), as shown for example in Fig. 3, therefore correspond to the trimer state $|\phi_{+}\rangle$ sitting above the upper band, the trimer state $|\phi_{-}\rangle$ sitting below the lower band and the end sites $|-m\rangle$ and $|m\rangle$ along with the third trimer state 
	$|\phi_{0}\rangle$ giving a triply degenerate level at zero energy between the bands. The band levels themselves are again even or odd superpositions of the positive and negative $i$ band states (as for case (a)) except that here each band contains $(m-2)$ levels (instead of the $m$ for case (a)). The total level count is thus still $2m+1$.
	
	These discussions of the bands and eigenstates for cases (a) and (b) show how energy gap protection arises. Choice of a sizeable coupling ratio (in our case we are taking $\Delta/\delta=40$) leads to gaps between the localised states and the bands of order $\Delta$. If the coupling ratio is lowered then the spread of the bands increases, relative to the separations between the localised states, reducing the effect of energy gap protection.

\subsection*{The Nature of the Localised States}

Further understanding of the nature of the localised states can arise by first considering the (extended) dimer region states. Note first that for our case (a), with positive site labels an odd site is strongly coupled to the next higher even site, whereas for negative site labels an even site is coupled strongly to the next higher odd site. We term the latter the A dimer configuration and the former the B dimer configuration, and clearly one maps to the other through interchange of the strong and weak couplings. It is well understood that these configurations are topologically distinct \cite{poli2014,bloch2013,schomerus2013,delplace2011}. This can be seen (following \cite{poli2014,schomerus2013}) by taking matrix elements for a dimer section of chain described by the Hamiltonian (\ref{hami}) (with $\epsilon_{i} = 0$) and writing the time-independent Schr\"odinger equation of the amplitudes for the excitation to be at site $n$
\begin{equation}
\nu \psi_{n} = J_{n,n+1} \psi_{n+1} + J_{n-1,n} \psi_{n-1} \; .
\end{equation}
For even $n$, dimer configuration A arises for $J_{n,n+1} = \Delta$ and $J_{n-1,n} = \delta$, and dimer configuration B arises for $J_{n,n+1} = \delta$ and $J_{n-1,n} = \Delta$. By grouping the site amplitudes into pairs for the $n$-th dimer, so $\underline{u}_{n} = (\psi_{2n},\psi_{2n+1})$ , these dimer states can be decomposed into Bloch waves as a function of a quasi-momentum $k$,
\begin{equation}
\underline{u}_{n} = \underline{u}(k) \exp(i k n) \; . 
\end{equation}
The two component vector $\underline{u}(k)$ can be viewed as the state of a pseudospin and it can be used to distinguish topologically between the A and B dimer configurations \cite{poli2014,schomerus2013,bloch2013,delplace2011}. 
One way to do this is to consider the polarisation vector of the pseudospin $(\langle \sigma_{x} \rangle,\langle \sigma_{y} \rangle,\langle \sigma_{z} \rangle)$. For all the extended dimer states this vector is confined to the $x-y$ plane. As $k$ varies over the Brillouin zone to map out the dimer energy band(s), this vector traces out a path in the $x-y$ plane which doesn't encircle the origin for dimer configuration A but does encircle it for dimer configuration B \cite{poli2014,schomerus2013,delplace2011}. Thus there is a clear topological (winding number) distinction between the configurations A and B. Another approach to identifying the distinction is through use of the Zak phase \cite{zak1989}. Whilst not gauge invariant, independently of the gauge used it is the same for all the states within a given configuration and its difference between states for configuration A and those for B is $\pi$, again demonstrating a topological distinction between the configurations \cite{bloch2013,delplace2011}.
Given all this, it is clear that our case (a) contains a single interface between two topologically distinct dimer chain regions. As seen in Figs.\ref{eigen} and \ref{energy}, at such an interface a single localised state arises, which is protected because it sits in a band gap that originates because two topologically distinct configurations meet. In the pseudospin language, this interface state can be viewed as having a polarisation in the $z$ direction and an imaginary quasi-momentum, giving rise to exponential localisation as a function of distance away from the interface \cite{poli2014,schomerus2013}.
The occurrence of a single interface state in case (a) is to be contrasted with the creation of a pair of interface states at the ends of a dimer chain when the couplings in the chains have their roles exchanged to create a defect at each end of the chain, such as in the work of reference \cite{delplace2011} where a continuous variation of the coupling ratio is performed. This paired state creation is effectively what happens in our case (b). Here three localised states arise due to the insertion of a "trimer defect" between two topologically distinct dimer configurations, along with a pair of localised states---one at each end of the system due to the introduction of defects at both ends.

To further understand the nature of the localised states already presented, these can be contrasted with localised states that do not arise because of an interface between two topologically distinct regions of chain. An example of the latter is a weakly coupled defect between two sections of uniformly coupled monomer chain. Whilst superficially similar, it is important to note the extended nature of the tails of the non-topological state: in the pseudospin framework discussed earlier, whilst the polarisation vector for this localised state is still in the $z$-direction, it does not have an imaginary quasi-momentum and so does not exhibit exponential localisation \cite{poli2014}. In addition, this localised state does not sit in an energy gap \cite{poli2014}. It is therefore not as  well protected against noise as the localised state in the equivalent topological example. 

\subsection*{Robustness Against Disorder}
 
	\begin{figure}
		\centering
		\includegraphics[scale=0.38]{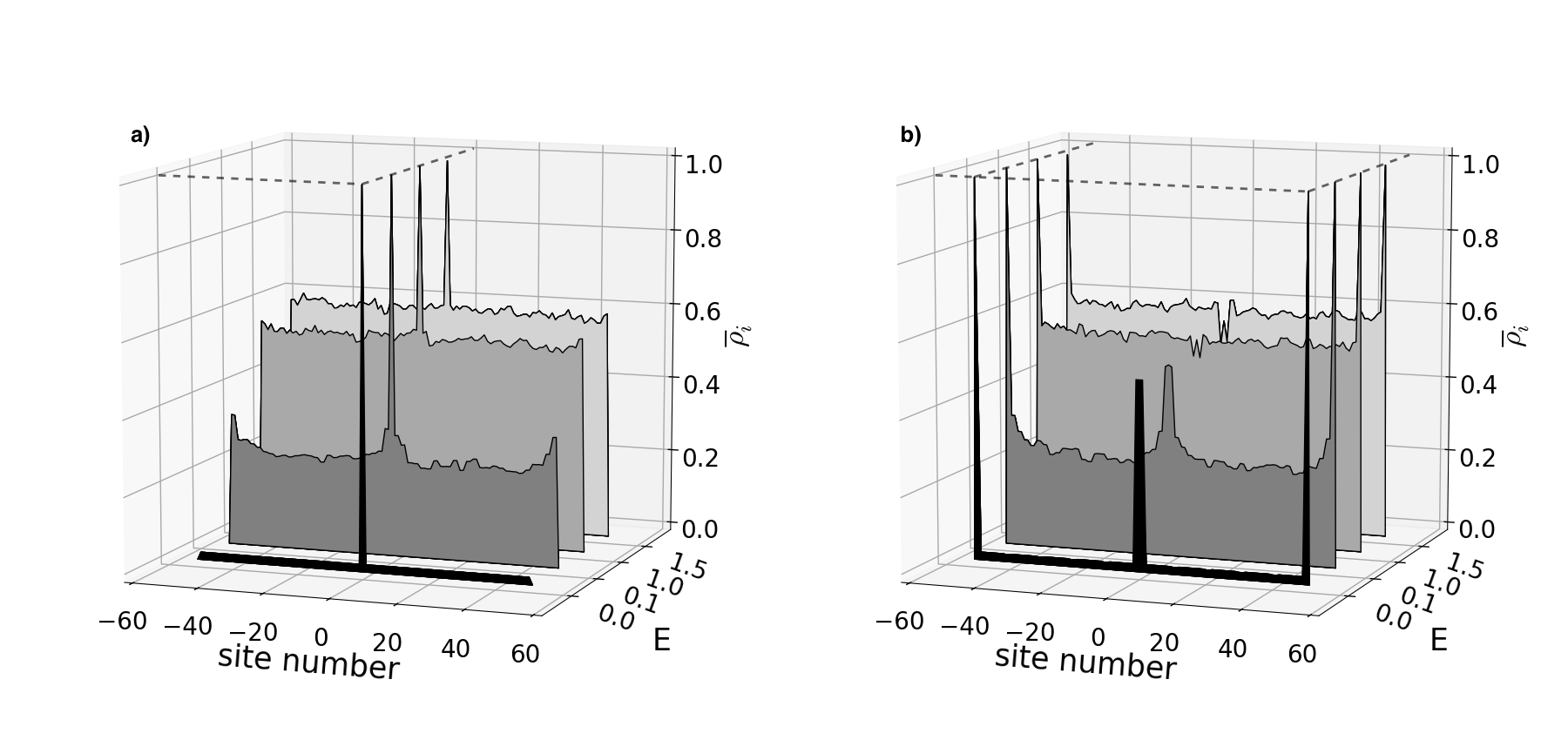}
		\caption{Maximum occupation probabilities for each site, averaged over 100 independent noise realisations, with $E=0.0$ (black), $E=0.1$ (darkgray), $E=1.0$ (gray) and $E=1.5$ (light gray) versus site number. For clarity, the y-axis representing the scale of the disorder $E$ has some band gap regions omitted; hence it is not to scale.}
		\label{noise}
	\end{figure}
	
    \begin{figure}
		\centering
		\includegraphics[scale=0.4]{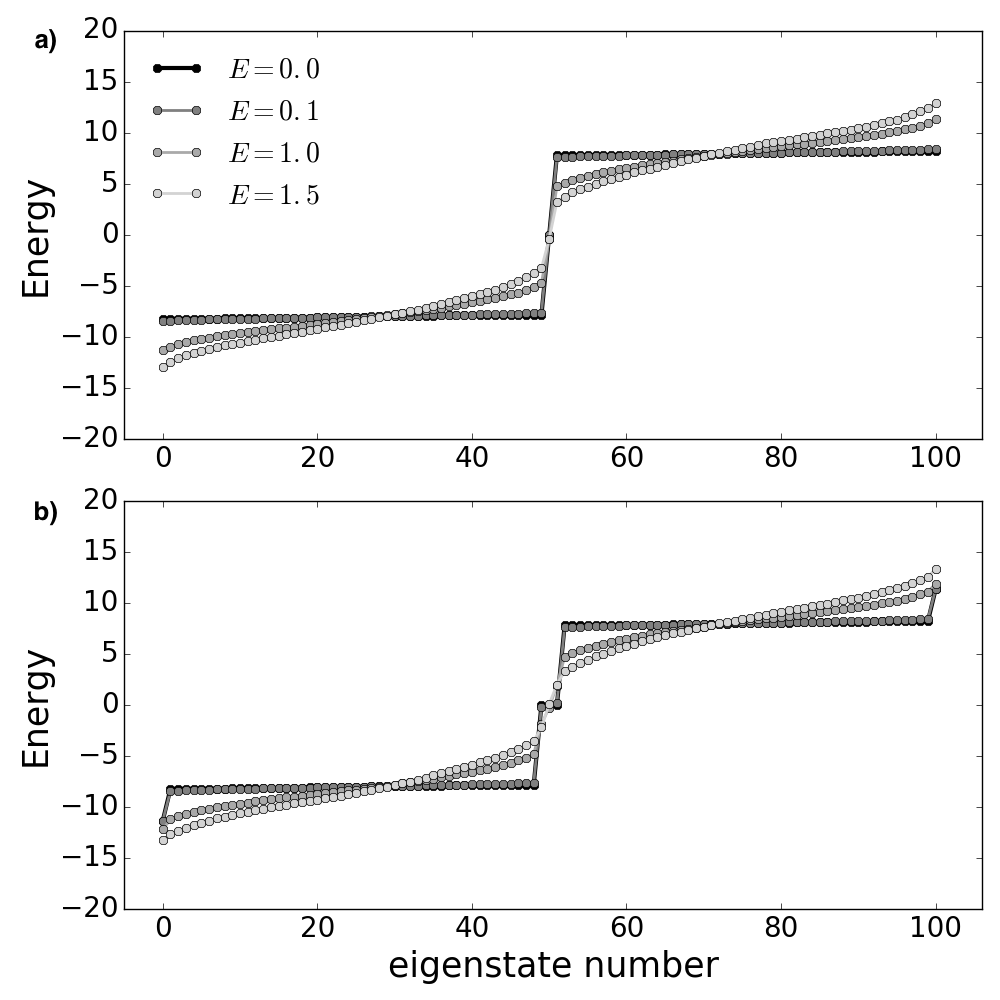}
		\caption{Energy spectrum of $N$=101 spin chains with site 0 (a) weakly and (b) strongly coupled, averaged over 100 independent noise realisations, and for different levels of disorder.  }
		\label{spectrum}
	\end{figure} 
    
For practical applications it is important to assess the robustness of these topologically localised spin chain states against fabrication defects. One approach to modelling such defects is to add random diagonal disorder to the Hamiltonian, which encompasses both the case of local fabrication defects and local fields fluctuations (most probable source of decoherence for the single excitation sector) \cite{ronke2011_1}. We set $\epsilon_{i} = E J_{s} d_{i}$, where $d_{i}$ is a random number from a uniform distribution between -0.5 and 0.5, and $E$ is a dimensionless parameter that sets the scale of the disorder. Due to the stochastic nature of these calculations, we average the maximum site occupancy probability for all the $N$ eigenstates over 100 realisations of the disorder (average denoted by a bar),
	
	\begin{equation}
	\bar{\rho_{i}} \equiv{\overline{\max_{n}|\langle i | \varphi_{n} \rangle|^{2}}}.
	\label{avg}
	\end{equation} 
	
	In Fig.\ref{noise} we present the maximum occupancy probabilities over the chain for cases with $E=0.1$, $E=1.0$, $E=1.5$ and without added disorder, $E=0.0$. For case (a) (left panel) it is seen that the probability of the state being in the middle of the chain remains peaked at unity, independent of the level of disorder. 
For case (b) (right panel) this protected behaviour is also observed for the states peaking at the ends sites and at sites $i=\pm 1$ (the eigenstate that is closely approximated by $|\phi_{0}\rangle$ of Eq.(\ref{eigenstates})). 
	Remarkably, this state is protected even when the growing Anderson localisation induced by increasing disorder \cite{anderson1958} is greater than the topologically-induced localization.
We note that for both chains the states protected by the gap around $E=0$ remain protected even when the overall noise amplitude is {\it larger} than the energy gap, up to $E\sim 3$, see also comments below. 
All of these points are a clear signature of the presence of  topologically protected localised states in the middle of the gap.

	Note also that the two states in Fig. \ref{noise} (with energies $\pm \sqrt{2}\Delta$) peaking at the mid point of the chain for case (b), the eigenstates that are closely approximated by states $|\phi_{-}\rangle$ and $|\phi_{+}\rangle$ of Eq.(\ref{eigenstates}), have some protection but start to be affected for $E \sim 1$. 
    
	All this behaviour can be further understood by observing the averaged energy spectrum for these same levels of disorder (Fig.\ref{spectrum}). With increasing disorder, the band energies spread and the band gaps shrink. Further increasing the disorder strength eventually closes the gaps and protection suffers. However, we have observed that the localisation remains very strong---even with a disorder strength of $E=3.0$ (three times the strong coupling value $\Delta$) the unit peaks decrease by only 4\%, so are still strongly localised with a probability $\rho_{i}\approx 0.96$. As already noted, the eigenstates approximated by states $|\phi_{-}\rangle$ and $|\phi_{+}\rangle$ of Eq.(\ref{eigenstates}) are less protected against disorder than the other localised states because their energy gaps are somewhat smaller than those of the zero energy states. Our results allow us also to comment on the temperature resilience of our system, which is important when applying a physical implementation of such a model to quantum information tasks. We note that it is the dimerised nature of the spin chains which is responsible for the characteristic that their spectra comprise two bulk energy bands (of extended states) along with various localised states. Thus, as long as in any physical implementation the variation of temperature (or indeed fabrication defects in the couplings) is smaller than the difference between weak and strong coupling energies, the dimerisation is preserved. Therefore, a gap between bands would persist, along with the presence of localised protected state(s).

\begin{figure}
	\centering
	\includegraphics[scale=0.4]{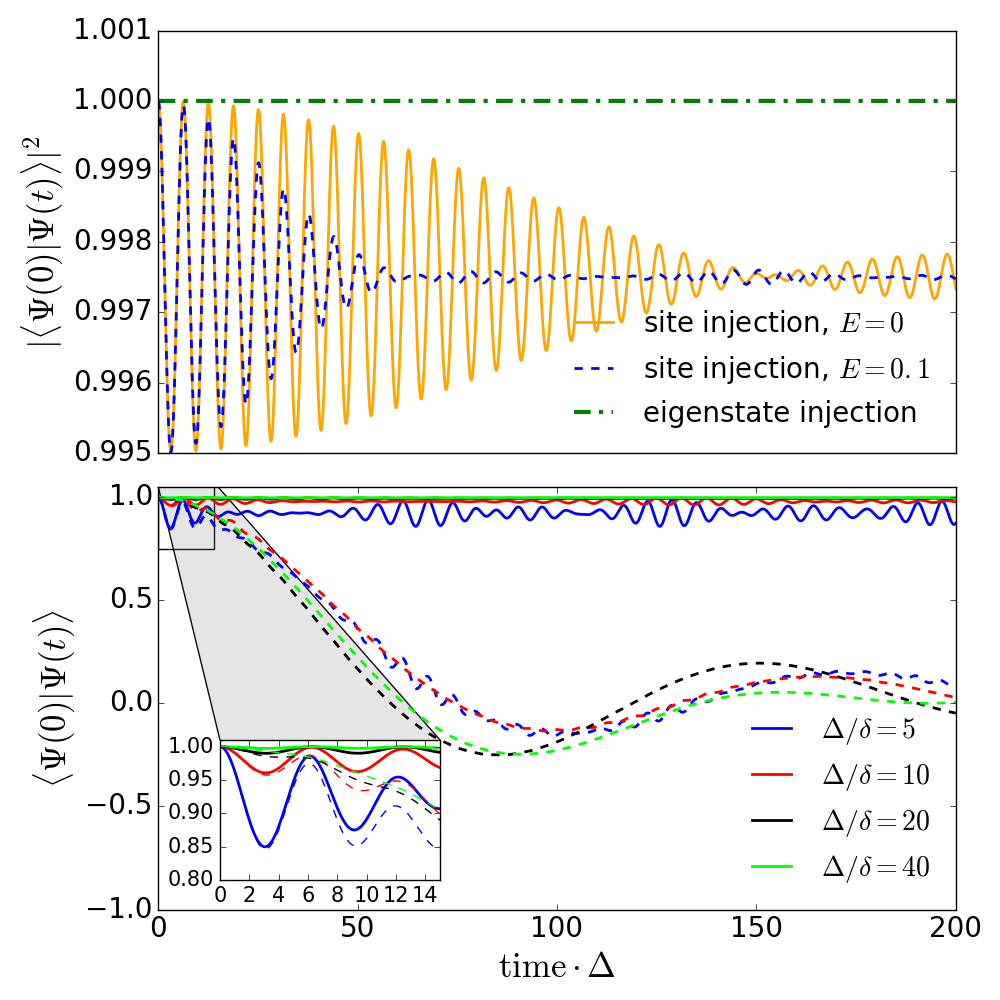}
	\caption{In the upper panel, fidelity of one initial excitation injected at the middle site (site encoding) -green- or localised eigenstate (eigenstate encoding) unperturbed -orange- or with $E=0.1$ level of disorder -dashed blue- of a case (a) $N$=21 chain and $\Delta/\delta=40$. In the lower panel, unperturbed phase dynamics of the excitation injected at the middle site (solid lines) and averaged disordered ($E$=0.1) phase dynamics over 100 realisations (dashed lines) for different coupling ratios ($\Delta/\delta$) of a case (a) $N$=21 chain and site encoding.   }
	\label{dynamics}
\end{figure}

\section*{Applications to Quantum Information}

In this section we wish to briefly explore applications to quantum information for the type of states analysed in the previous section.

\subsection*{Encoding}

Encoding of quantum information in a spin chain with single-spin occupancy in each site and at half-filling (the case at hand) is usually done by direct 
correspondence between a single site/spin and a single qubit, see e.g. \cite{spiller2007,ronke2011_2}. In this case a spin chain of $N$ sites corresponds to a chain of $N$ qubits, so a 4-spin chain 
in the state $|\uparrow,\downarrow,\downarrow,\uparrow\rangle$ would correspond to the quantum register (or quantum memory) $|1,0,0,1\rangle$. The chain can then be used as a quantum memory if its 
only function is to store quantum information, or used as quantum register if one wishes (or is able) to use the interactions between the spins in order to perform 
controlled quantum dynamics. As an example, in Ref \cite{ronke2011_2} spin-spin interactions are used to knit chains of 4-qubit states for one-way quantum computation. From now on we will call this encoding `site encoding'. Suppose that we inject into a system that has all spins/sites down, $|0\rangle=|00..0\rangle$, an arbitrary qubit state of $\alpha |0\rangle + \beta |1\rangle$ into the site $s$ which is the dominant site of a localised eigenstate. The subsequent dependence of the state is,

\begin{equation}
\label{siteencoding}
	|\Psi_{s}(t)\rangle=\alpha |0\rangle + \beta \sum_{n}\langle \varphi_{n}|00..1_{s}..00\rangle e^{-iE_{n}t/\hbar}|\varphi_{n}\rangle
\end{equation} 

This time dependence arises because, although the sum will have a dominant contribution from the localised state, we are injecting into a superposition of all the Hamiltonian eigenstates ($|\varphi_{n}\rangle$) that have non-zero amplitude at site $s$.

In the case at hand, because of the potentially very strong spatial localisation of the protected states, we can either use the encoding described above, or 
encode instead a single logical qubit for each of the topologically localised eigenstates ($|\varphi_{L}\rangle$), with presence or absence of the excitation being the $|1_{L}\rangle$ or $|0_{L}\rangle$ qubit state, respectively, such that,

\begin{equation}
\label{stateencoding}
|\Psi_{L}(t)\rangle=\alpha |0_{L}\rangle + \beta e^{-iE_{L}t/\hbar}|1_{L}\rangle
\end{equation} 

This latter `eigenstate encoding' allows  to use these protected states as qubits even when their spatial localisation is not so strong and the physical states extend over several spins, as long as the eigenstate remains isolated in an energy gap and thus accessible through energy-specific addressing. In this case the number of logical qubits that a chain could host would depend on the number of topological solitons embedded in its couplings' structure.  

Injecting and retrieving information from a spin chain will depend on the specific hardware used for embedding the mathematical concept of  spin chain. In Refs. \cite{spiller2007} and \cite{ronke2011_1}, for example, a spin chain was embedded in a chain of self-assembled quantum dots. In this specific case injecting or retrieving qubits from the chain can be done using trains of laser pulses \cite{spiller2007,ronke2011_1}.

\begin{figure}
	\centering
	\includegraphics[scale=0.35]{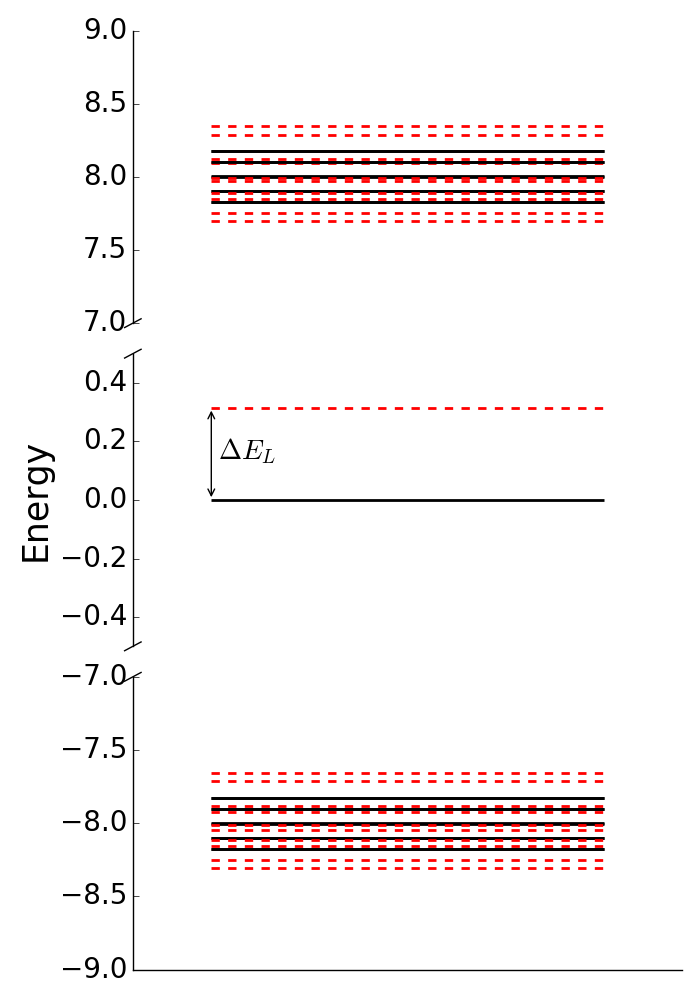}
	\caption{Energy splitting between the unperturbed (black lines) and one preturbed realisation with $E=0.1$ (dashed) of the states of a case (a) $N$=21 chain. We note that for a single realisation the energy of the localised state can shift either up or down but will remain in the gap (see averaged case shown in Fig.\ref{spectrum}). }
	\label{quantumMem}
\end{figure}

\begin{figure}
	\centering
	\includegraphics[scale=0.4]{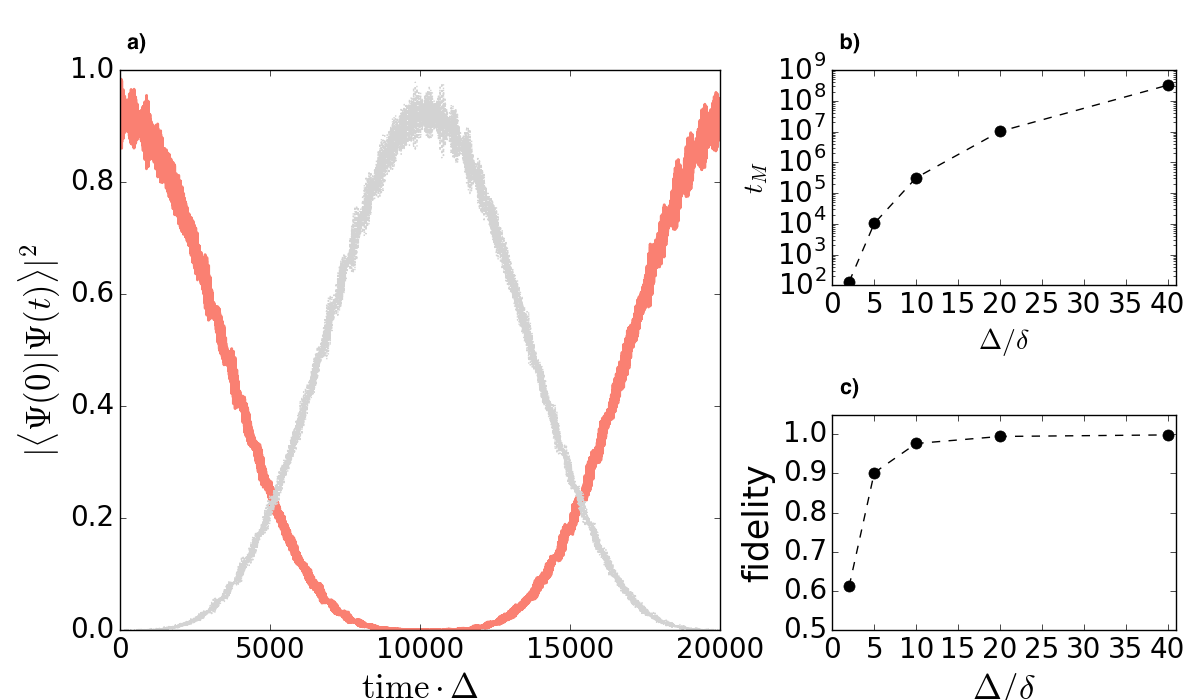}
	\caption{(a) Fidelity of an initial excitation injected at one end (site encoding) of a case (b) $N$=21 chain with $\Delta/\delta=5$ against its initial state $|10...0\rangle$ (red line) and mirroring state $|0...01\rangle$ (gray line), (b) mirroring time ($t_M$) for different coupling ratios and $N$=21 and (c) fidelity at the mirroring time for different coupling ratios and $N$=21.}
	\label{pst}
\end{figure}

\subsection*{Quantum Memory}

The time evolution of an initial state injected at any of the zero-energy non-overlapping localised states or at their corresponding peaking sites will display a high fidelity ($F=|\langle\Psi(0)|\Psi(t)\rangle|^2$) of finding the state in its initial position, as shown at the upper panel of Fig. \ref{dynamics}. The impact of the encoding chosen can be clearly seen here: while the fidelity for an injected excitation at the localised site (site encoding) oscillates around F=0.997, the probability of finding the state initially injected at the localised state (eigenstate encoding) is constant at unity. The oscillations in the site encoding fidelity are due to small components of non-localised energy eigenstates (Eq.\ref{siteencoding}), so the frequency is dominated by $\Delta$. These contributions from non-localised energy eigenstates become more significant as the coupling ratio is reduced. 

For both encodings, due to the protected nature of the zero-energy states the amplitude of the dynamics remains protected against disorder as the state remains within the energy gap (as seen on the upper panel of Fig.\ref{dynamics} for the injection at site $i=0$ and perturbation $E=0.1$). However, when looking at the averaged phase (lower panel of Fig.\ref{dynamics}) for the site encoding we observe a phase error when the disorder added is of a strength $E$=0.1 which is similar for all the coupling ratios. 

This phase error makes our site encoding on the system presented unsuitable for quantum memory applications, as it represents a relative phase on a qubit superposition involving an amplitude with no excitation and an amplitude with one site excitation (see Eq.\ref{siteencoding}). 

Thus the eigenstate encoding alternative here becomes more relevant. For case (a), even though the localised state will remain protected (within an energy gap) when disorder is added, it will shift an energy $\Delta E_{L}$ (see Fig.\ref{quantumMem}). In case of disorder then the qubit should be encoded into the perturbed localised eigenstate at energy $\Delta E_{L}$. As the state remains well within the large gap, its energy should be relatively easily identified experimentally. Hence although both encodings will present a phase oscillation, the use of the eigenstate encoding and the fact that we will be injecting into a well known eigenstate at $\Delta E_{L}$ will allow us to know the periodicity of this phase and hence correct for it, if needed. 

\subsection*{Perfect State Transfer of Quantum Information}

As already mentioned, one additional particularity of some spin chains is the possibility of tuning them to allow for perfect state transfer (PST), a property with already well-known applications in quantum information processing \cite{kay2010,bose2007,spiller2007}. For our system, we will show that this can be achieved by manipulating the chain properties (either length and/or coupling ratio) such that two localised states overlap. We achieve this here by decreasing the length of our (b) type chain to 21 sites and setting the coupling ratio to $\Delta/\delta=5$.

As shown in panel (a) of Fig.\ref{pst}, and due to the superposed nature of the two now symmetric and antisymmetric localised states, we observe that the state initially injected at one of the edge sites of the chain transfers to its opposite (mirror) site at the other edge of the chain at a $time\simeq 1.0\cdot10^4$ (mirroring time, $t_M$) in inverse $\Delta$ units and its fidelity revives when the state comes back to its initial site after twice this time. Given that the transfer time will depend on the overlap (and thus energy splitting) between the localised states and at the expense of taking longer times, the fidelity can be increased by increasing the coupling ratio $\Delta/\delta$, as shown in panels (b) and (c) of Fig.\ref{pst}. Exploration of these dynamics, along with application of these effects to quantum gates and information transfer, will form the subjects of further studies.

\section*{Conclusions}
	
	In this paper we have investigated the presence and robustness of topologically localised states in engineered spin chains, inspired by the SSH model. The presence of these states can be selectively manipulated through control of the chain coupling distribution and length. Localised states can be engineered to exist at the centre and/or the ends of the chains [as in cases (a) and (b)]. Other chain arrangements can generate different localisation patterns; these will be reported in a further work.
	
	We have shown that these topologically localised states exhibit a high level of protection to increasing disorder, with higher protection resulting from larger energy gaps. 
	Such topologically localised states are to be contrasted with those that exhibit growing Anderson localisation \cite{anderson1958} with increasing disorder. It can be seen from Fig.\ref{noise} that the former persist with high fidelity even when other states are localising to other sites with significant probability. Such robustness is an important requirement when thinking of quantum architecture components. The presence of this property provides an interesting system presenting two or more degenerate states where to encode topologically protected quantum information \cite{kitaev2003}. It may thus be a good candidate from which to design a quantum memory device \cite{dennis2002}. 
	
	Members of these families of spin chains, when engineered and combined, could also represent promising elements for the construction of more complex quantum logic networks \cite{ronke2011_2}, thus providing a novel system with which to perform quantum information processing. In order to investigate such applications, future work will examine the dynamics, state transport and computational abilities of appropriate spin chain systems in a more extensive way, along with the robustness of these  applications to disorder.

\bibliography{papers_ire}

\section*{Author contributions statement}

M.P.E. did the calculations. All the authors interpreted the data and wrote and reviewed the manuscript.

\section*{Additional information}

The authors declare no competing financial interests.

\end{document}